\documentclass{elsarticle}

\usepackage{graphicx}% Include figure files
\usepackage{amsmath}

\begin{document}

\title{What makes a phase transition? Analysis of the random satisfiability problem}

\author[tszk]{Katharina A. Zweig}
\ead{nina@ninasnet.de}
\author[kcsp]{Gergely Palla\corref{cor1}}
\ead{pallag@hal.elte.hu}
\author[tszk,kcsp]{Tam{\'a}s Vicsek}
\ead{vicsek@hal.elte.hu} 
\cortext[cor1]{Corresponding author, tel:+36-1-3722768, fax:+36-1-3722757}

\address[tszk]{Department of Biological Physics, E{\"o}tv{\"o}s University, 1117 P\'azm\'any P. Stny 1/A, Budapest, Hungary}
\address[kcsp]{Statistical and Biological Physics Research Group of HAS, E\"otv\"os University, 1117 P\'azm\'any P. Stny 1/A, Budapest, Hungary}
%\ead{\mailto{nina@ninasnet.de}, \mailto{pallag@angel.elte.hu}, \mailto{vicsek@angel.elte.de}}

\begin{abstract}
In the last $30$ years it was found that many combinatorial systems undergo 
phase transitions. 
One of the most important examples of these can be found among the 
random $k$-satisfiability problems (often referred to as $k$-SAT), 
asking whether there exists an assignment 
of Boolean values satisfying a Boolean formula composed of clauses with 
$k$ random variables each. The random 3-SAT problem is reported to
show various phase transitions at different critical values of the ratio 
of the number of clauses to the number of variables. The most famous of
 these occurs when the probability 
of finding a satisfiable instance suddenly drops from 1 to 0. This
 transition is associated with a rise in the 
hardness of the problem, but until now the correlation between any of the 
proposed phase transitions and the hardness is not totally clear. In this paper
 we will first show numerically that the number of solutions universally 
follows a lognormal distribution, thereby explaining the puzzling question of 
why the number of solutions is still exponential at the critical point.
Moreover we 
provide evidence that the hardness of the closely related problem of counting 
the total number of solutions does not show any phase transition-like 
behavior. This raises the question of whether the probability of finding
a satisfiable instance is really an order 
parameter of a phase transition or whether it is more likely to just show a 
simple sharp threshold phenomenon. More generally, this paper aims at starting 
a discussion where a simple sharp threshold phenomenon turns into a genuine 
phase transition.\\
\end{abstract}


\maketitle

\section{Introduction}

The analysis of phase transitions and the associated microscopic structures is a well-developed scientific approach in physics. In real systems, the observation of phases and their different macroscopic behavior comes first, and a subsequent analysis reveals how the structure of one phase is transformed into the structure of the other phase. This transition is associated with the change of a so-called {\it control parameter}, such as the temperature. Most interesting are abrupt changes in functions measuring the macroscopic behavior, e.g., the density or heat capacity, that happen with small changes in the control parameter. The function showing the non-analytic behavior or {\it singularities} is the {\it order parameter} of the system and can be seen as a fingerprint of the underlying phase transition. Starting with the analysis of random graphs \cite{Bollobas2001} and simple percolation models \cite{Stauffer1994,Bollobas2006}, combinatorial objects came into the focus of statistical physicists. A thorough analysis revealed that these simple systems also show phase transitions. 

Whereas in percolating systems the phases and their different behaviors are visually accessible, this is not the case for other combinatorial systems with a proposed phase transition. One of the most important of these systems is the so-called {\it satisfiability problem (SAT)}. Given some Boolean formula, it asks whether there exists an assignment of Boolean values to its variables such that it is satisfied, i.e., such that it evaluates to $true$. SAT problems belong to the set of NP-hard problems, i.e., so far there is no algorithm to solve them in polynomial time \cite{Garey1979}. As with many other NP-hard problems, satisfiability problems arise not only in theory but also in industry, e.g., in automotive configuration \cite{Sinz2003}, in software and hardware design \cite{k-itfhv-98}, biological sciences \cite{Elser2006}, and artificial intelligence \cite{Biere2009}. 
Since satisfiability problems are so abundant, understanding when and why they are hard and developing better algorithms is crucial.
A classic family for analyzing the hardness is the random $k$-SAT family in which the $k$ variables of each clause are drawn uniformly at random and without repetition from the set of all variables. Each variable is negated with probability $0.5$. The ratio between the number of clauses $m$ and variables $n$ denoted by $\alpha\equiv m/n$ parameterizes the probability P[UNSAT] of finding an unsatisfiable instance at a given  $\alpha$. It was observed early \cite{Cheeseman1991,Mitchell1992} that plotting P[UNSAT] against $\alpha$ shows a sharp threshold behavior at some critical $\alpha_c$. Furthermore, around this $\alpha_c$ it also takes various algorithms the longest time to solve random 3-SAT problems, i.e., the problems are {\it hard}. 
To quantify the hardness, either the number of distinct steps of the solving algorithm is counted, or simply the time measured until the problem is solved. The divergence of the hardness together with the sudden jump of P[UNSAT] at some critical $\alpha$ resembles a phase transition-like behavior \cite{Hartmann2005,Percus2006}. The numerical analysis of this sharp threshold behavior resulted in $\alpha_c = 4.15\pm 0.05$  \cite{Kirkpatrick1994}.
Kirkpatrick and Selman could also show that there is a non-trivial finite size effect, i.e., that the width of the window in which the transition takes place is proportional to $n^{-2/3}$ for $3$-SAT. It is thought that this sharp threshold phenomenon is of first order, i.e., in the limit of infinite system size and for $\alpha < \alpha_c$ P[UNSAT] $= 0$, and for $\alpha > \alpha_c$ P[UNSAT] $= 1$. For 2-SAT, this could be rigorously shown \cite{Chvatal1992}, but for all $k \geq 3$ it is an open question. Note that $2$-SAT itself is not NP-hard \cite{Garey1979}.
To analyze the nature of this sharp threshold behavior, the $k-SAT$ problem can also be represented as a spin-glass model, and different theoretical analyzes have arisen from this approach \cite{Monasson1999,Mezard2002b,Mezard2002,Krakala2007}. 
Since these theoretical analyzes rely on the thermodynamical limit whereas numerical approaches can only tackle system sizes of up to $100$ or even be restricted to system sizes below $40$ (depending on the specific question), it is not surprising that none of the theoretical approaches matches the numerical value of $\alpha_c = 4.15\pm 0.05$. The approach that comes closest is based on the analysis of survey propagation that results in $\alpha_c=4.267$, which is believed to be exact \cite{Mezard2002}. The applied order parameter is very technical and it is difficult to analyze how it relates to P[UNSAT]. 

To find out more about the behavior of $3$-SAT, we first repeated the experiment of Kirkpatrick and Selman, and increased the then available system size from $100$ to $200$. A subsequent finite size scaling is much more in accordance with the old value of $\alpha_c=4.15\pm 0.05$ than with $\alpha_c=4.267$. In the second step, we aim at understanding a different parameter, namely the {\it entropy} of the system, i.e., the logarithm of the number of solutions a satisfiable instance has. It was shown by an approach from statistical physics that the entropy is still finite  at $\alpha_c$, i.e., the number of solutions is still exponential \cite{Monasson1996}. Monasson and Zecchina state that `` hence (...) the transition itself is due to the abrupt appearance of logical contradictions in all solutions and not to the progressive decreasing of the number of these solutions down to zero.'' Such a sudden emergence of logical contradictions on a macroscopic level would be a good sign of a genuine phase transition. 

In this paper we give numerical evidence that the explanation for the finite entropy at $\alpha_c$ is far simpler, namely that the average number of solutions of satisfiable instances is universally described by a lognormal distribution over a range of different system sizes and $4.0 \leq \alpha \leq 4.5$. This means that, although many of the instances are already unsatisfied at $\alpha_c$, some of the satisfiable instances have a large number of solutions left, which accounts for the high average number of solutions. A lognormal distribution can be the result of the iterative application of a factor drawn from some distribution. This raises the question of whether the phase transition of P[UNSAT] may be only a sharp threshold phenomenon that is not based on the non-trivial restructuring of interacting entities. In the following we will first discuss our numerical findings regarding the average number of solutions, then give an alternative explanation for the rise of the hardness at $\alpha_c$ and finally discuss some simple models with different kinds of sharp threshold phenomena. The last model shows qualitatively the same behavior as P[UNSAT]. 

In summary, we do not attack the idea that $k$-SAT shows phase transitions in general but we put on display some simple explanations and models that raise doubt about whether the proposed phase transition of P[UNSAT] is more than a simple sharp threshold phenomenon. In general, the once obvious border between first order and continuous phase transitions and their respective properties has become so blurred that scientists from neighboring disciplines, e.g., computer scientists or chemists and even statistical physicists not specialized in spin-glasses, have difficulties to find out what are the properties that define a phase transition. Our main contribution in this paper are thus the above mentioned toy models that are so simple that they cannot be considered to have a genuine phase transition. Still, they mimic some important properties of the $3$-SAT system. With this we would like to open a discussion with the spin-glass community to understand what differentiates the simple models from $3$-SAT and what exactly makes a phase transition. The paper thus aims at starting a discussion of the difference between a mere sharp threshold phenomenon and a genuine phase transition. We hope that a discussion of what properties are required for acknowledging a phase transition will help to support the interdisciplinary discussion in this area.

The paper is organized as follows: After giving some definitions in Sec. \ref{definitions}, we will discuss in Sec. \ref{sat} the question of whether the sharp threshold phenomenon of P[UNSAT] is directly caused by a continuous phase transition of an order parameter related to P[UNSAT]. We will furthermore discuss whether there is any evidence at all for the existence of two different phases. Sec. \ref{models} finally introduces two simple statistical models that show similar phase transition-like behavior without any underlying interacting elements. The first one is clearly trivial, while the second shows a non-trivial finite size scaling effect. From these models, we develop a simple toy model that qualitatively shows the same properties as P[UNSAT] in random $3$-SAT. Finally, we discuss our findings in Sec. \ref{summary}.

\section{Definitions}
\label{definitions}
Let $V$ be a set of $n$ variables $\{v_1, \dots, v_n\}$. Each variable has two literals, a positive literal denoted by $v_i$ and a negated literal denoted by $-v_i$. A Boolean formula in {\it conjunctive normal form (CNF)} consists of $m$ subsets of {\bf and}-connected literals, called {\it clauses} or {\it constraints}. The clauses are {\bf or}-connected. An {\it assignment} is a function $a: V \rightarrow \{true,\ false\}^n$ that assigns each variable a Boolean value, i.e., {\it true} or {\it false}. With a given Boolean formula in CNF and a given assignment, the formula can be {\it evaluated}: a positive literal which is assigned $true$ evaluates to $true$, and to $false$ if it is assigned $false$. A negated literal which is assigned $false$ evaluates to $true$ and to $false$ otherwise. A {\it clause} evaluates to $true$ if at least one of its literals evaluates to $true$, and the whole formula evaluates to $true$ if all clauses evaluate to $true$. The {\it satisfiability problem}, or SAT problem for short, asks whether a given Boolean formula has at least one assignment such that it evaluates to $true$. Such an instance is called {\it satisfiable (sat)}, and one where no satisfying assignment can be found is called {\it unsatisfiable (unsat)}. If all clauses contain $k$ literals, we speak of $k$-SAT. If, moreover, the instance is created by choosing the $k$ literals uniformly at random without repetition, we speak of random $k$-SAT. $\alpha$ denotes the ratio between the number of clauses $m$ and the number of variables $n$ in a random $k$-SAT instance. 

For each two assignments $a$ and $a'$, the {Hamming distance} $d(a, a')$ is defined as the number of different assignments to the variables.

The SAT problem can be solved by different algorithms, the most widely used being based on the following scheme, first proposed by Davis et al. \cite{Davis1962}. It is a kind of trial-and-error procedure in which a growing subset of variables is assigned Boolean values until we either find a solution or encounter a contradiction. In each step, take one of the variables that is yet unassigned and assign either $true$ or $false$ to it. Say, variable $v_i$ is assigned $true$. Now, the instance can be simplified by (temporarily) removing all clauses which contain the positive literal of $v_i$ since they are already satisfied. Furthermore, we can temporarily remove the negated literal from all clauses since it cannot contribute to the satisfaction of the clauses it is contained in. If after this step all clauses have been removed, we have found a solution to the problem. If we encounter an empty clause, all of its originally contained variables have been assigned the wrong value and thus we have found a contradiction. In this case, we have to backtrack and restore the instance up to the point where $v_i$ was unassigned. Then, the same procedure is tried, but assigning $false$ to $v_i$.  As long as there is no solution and no contradiction in the simplified instance, we simply proceed with the partial assignment. If all decisions lead to contradictions, the instance is unsatisfiable. There are many improvements to this basic scheme, e.g., specifying an order in which the variables are assigned \cite{Marques-Silva1999} and learning \cite{Marques-Silva1996}. One basic improvement is {\it unit propagation}, i.e., whenever a clause has only one literal left, it can only be satisfied when the variable's assignment is set accordingly. Note that the assignment of such a variable is called a {\it dependent decision} while the assignment of Boolean values to all other so-called {\it free} variables is called {\it independent decision}. 

\section{Random 3-SAT}
\label{sat}
It is well known that $3$-SAT belongs to the set of the so-called $NP$-hard problems, i.e., problems for which so far no algorithm with polynomial runtime has been found \cite{Garey1979}. In the worst case, finding a solution to these problems can take exponential time such that even relatively small instances cannot be solved within months. On the other hand, many real-world SAT problems can be solved in a short time despite their huge size. Since this behavior is not well understood, research has been dedicated to understanding why and how hard instances emerge and what their structure looks like. 

\begin{figure}
\centerline{\includegraphics[width=0.9\textwidth]{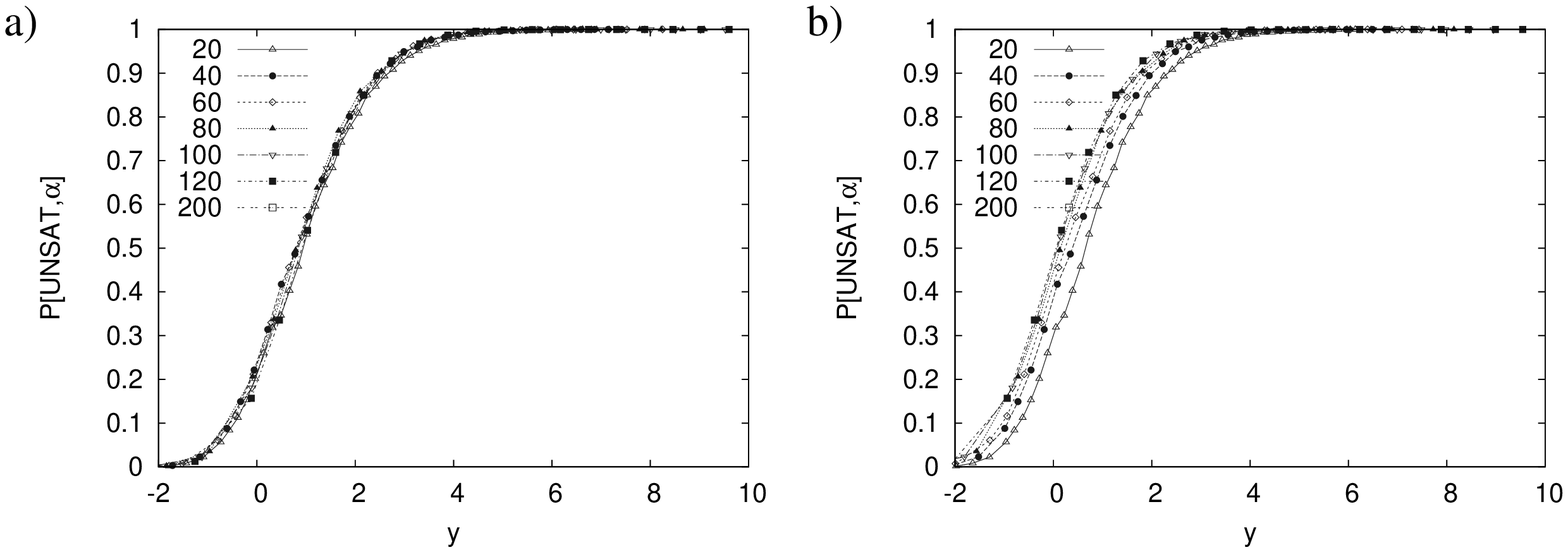}}
\caption{\label{finiteSize} a) Scaling of P[UNSAT] against $y = N^{0.66}(\alpha-4.12)/4.12$. 
b) Scaling of P[UNSAT] against $y = N^{0.66}(\alpha-4.267)/4.267$.
}
\end{figure}

It was observed early \cite{Mitchell1992, Cheeseman1991} that plotting P[UNSAT] against $\alpha$  shows a sudden jump at some  value $\alpha_c$ independent of the system size $n$. Furthermore, around this  value $\alpha_c$ it also takes various algorithms the longest time to solve random 3-SAT problems. This divergence of the hardness and the sudden jump of P[UNSAT] at some universal $\alpha$ resembles a phase transition-like behavior \cite{Hartmann2005,Percus2006}. In their classic paper from 1994, Kirkpatrick and Selman used the well-understood model of percolation in growing random graphs and the techniques deployed in this area for the identification of critical phenomena in random $3$-SAT: ``We use finite-size scaling, a method from statistical physics in which the observation of how the width of a transition narrows with increasing sample size gives direct evidence for critical behavior at a phase transition.'' They scaled the curves for different $k$ according to $n^{\nu}*(\alpha-\alpha_c)/\alpha_c$ and evaluated $\alpha_c$ to be $4.15\pm 0.05$ and the critical exponent $\nu$ for $k=3$ to be $2/3$ \cite{Kirkpatrick1994}. 
Today a value of $\alpha_c=4.267$ is often cited for the P[UNSAT] threshold \cite{Mezard2002}, but plotting P[UNSAT] against the rescaled parameter $y = n^{0.66}(\alpha-4.12)/4.12$ yields a much better scaling than that for the rescaled parameter $y=n^{0.66}(\alpha-4.267)/4.267$ (s. Figure~\ref{finiteSize}). The reason for this mismatch is not totally clear. It could be due to the still quite small system size in our experiments.

In this paper we suggest that the observed threshold phenomenon of P[UNSAT] is not so much a sign of criticality but simply caused by the law of large numbers. In general it is not easy to prove that an observed sharp threshold behavior is not caused by the critical behavior associated with a phase transition since there are many possible interactions that could be causing it. In the next section we will first analyze the typical number of solutions, which is closely related to the entropy of the system.

\subsection{Number of solutions}

A first-order phase transition is deeply connected to a sudden increase in {\it order}. For example, when water freezes the molecules are fitted into a neat structure that shows high order. It is difficult to see intuitively what kind of order is measured by P[UNSAT]. However, when a continuous phase transition is 
studied using an {\it existence parameter} instead of a 
{\it quantitative parameter}, it may seem to be rather like a 
first order transition, as we will exemplify in the case of site percolation in 2D.
%But sometimes a seemingly first-order phase transition is just a by-product of a second-order phase transition as we will exemplify in the case of site percolation in 2D. 
Here, one can ask about the behavior of two different but related parameters: ``Is there a biggest connected component (BCC) of size O(n)?'' (this is the existence parameter) or ``What is the size of the BCC ?'' (and this is the quantitative parameter). Plotting the relative size of the BCC shows a continuous phase transition at some critical value, i.e., the first parameter is a quantitative one that reveals the complex behavior of the system. At the critical value, a finite fraction of all vertices is spanned by the BCC, i.e., it has size $O(n)$. Since the second parameter just asks for the existence of a BCC with size $O(n)$, it will trivially show a first-order phase transition-like behavior at the same value \cite{Stauffer1994}. Thus, in this system, the seemingly first-order phase transition-like behavior of the existence parameter is just a trivial implication from the true continuous phase transition concerning the quantitative 
parameter.
Since P[UNSAT] asks whether there exists a solution or not, we first analyzed whether the seemingly first-order phase transition of P[UNSAT] also belongs to this type, i.e., whether it is an indicator of a more complex continuous phase transition of a related quantity like the behavior of the number of solutions.

\begin{figure}
\centerline{
\includegraphics[width = 0.6\textwidth]{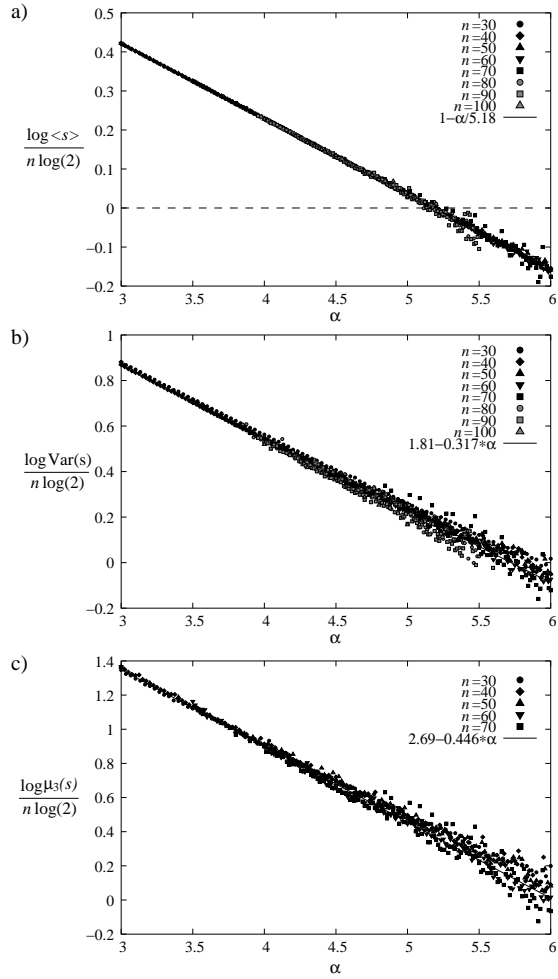}
}
\caption{\label{avgNumberSol} a) The rescaled average number of solutions as a function of
 $\alpha$, showing the behavior predicted by Eq. \ref{avgNumb}. Interestingly, the
 variance and the third central moment, $\mu_3$, follow similar rules
as shown in panels (b) and (c), respectively.}
\end{figure}

An instance is $unsat$ if and only if it has no solution---this is a typical {\it existence parameter}. A possible quantitative parameter of which this existence parameter could be an indicator is the average number of solutions. The logarithm of this quantity is the entropy of the system at a given $\alpha$ \cite{Monasson1996}.
Figure~\ref{avgNumberSol} shows that the average number of solutions $<s>$ can be fitted to a simple exponential law, i.e.,  
\begin{equation}
\label{avgNumb}
<s> = 2^n\left(\frac 7 8\right)^m. 
\end{equation}
This simple behavior of the average number of solutions coincides with the so-called annealed estimate of the number of solutions \cite{Kirkpatrick1994}, which is based on the fact that any solution will be `killed' with a probability of $1/8$ by a clause drawn uniformly at random. But although this estimate has been used for a long time, it is surprising that the average number of solutions follows it so closely since it does not take into account that in reality the solutions' probability to be deleted are dependent: i.e., two very similar solutions have a higher probability to be killed by the same constraint whereas two solutions that assign the opposite values to variables can never be killed by the same constraint. Thus, it is still surprising that the average number of solutions universally follows this simple law for all system sizes. Furthermore, Figure~\ref{avgNumberSol} reveals that at $\alpha_c$ there is---on average---still an exponential number of solutions although we know that the probability of finding a satisfiable instance drops to zero for large system sizes. This has also been proven rigorously by \cite{Monasson1996}. It is clear that without the gap between the critical value of $\alpha_c$ and the point $\alpha = 5.19$ where the {\it average} number of solutions becomes $1$, there would not have been much interest in the seemingly critical behavior of P[UNSAT]. 

\begin{figure}
\centerline{
\includegraphics[width = 0.6\textwidth]{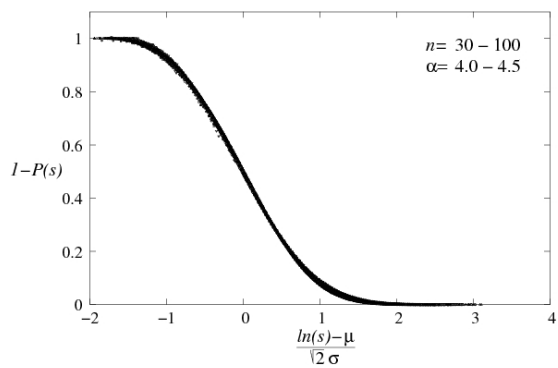}}
\caption{\label{distNumbSol} The rescaled cumulative distribution of the number of solutions of satisfiable instances over a large range of $\alpha=4.0-4.5$ and system sizes $n=30-100$. The $\mu$ and $\sigma$ denote the two fitting parameters of the lognormal distribution used for the rescaling.}
\end{figure}

The only possibility to achieve an exponential average number of solutions at $\alpha_c$ and P[UNSAT] $\rightarrow 0$ for $n \rightarrow \infty$ is to have a strongly right-skewed distribution of the number of solutions an instance has. Indeed, as Figure~\ref{distNumbSol} shows,  the distributions of satisfiable instances displays a universal behavior. Over an interval of $\alpha=4.0-4.5$ and different system sizes $n=30-100$, the cumulative distribution of the number of satisfiable instances, $P(s)$, can be fitted by the cumulative distribution of a lognormal distribution given as
\begin{equation}
 P(s)=\frac{1}{2}+\frac{1}{2}{\rm erf}\left[\frac{\ln(s)-\mu}{\sigma\sqrt{2}}\right],
\label{eq:lognormal}
\end{equation}
where $\mu$ and $\sigma$ correspond to the mean and the standard deviation 
of  $\ln(s)$, and erf denotes the error function.
%{\bf [Even if we could not get the delta-function out of this, do we know how $P[unsat]$ grows with $n$ at $\alpha_c$? Can we show that it goes to $1$, i.e., that the transition is sharp indeed?]}.

\begin{figure}
\centerline{
\includegraphics[width = 0.6\textwidth]{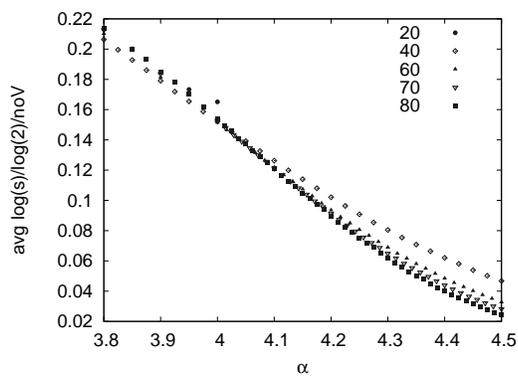}}
\caption{\label{quenchedAverage}The quenched average of the number of solutions.}
\end{figure}

The lognormal distribution of $s$ explains that there is no need of a sudden drop of $<s>$ at $\alpha_c$ since the average is dominated by some instances with a high number of solutions, although most instances are already unsatisfiable. In summary, neither the typical number of solutions nor its distribution shows critical behavior around $\alpha_c$.
Since we know now that the distribution of $s$ is highly skewed, another intuitive measure is the {\it quenched average}, i.e., the average $<\log(s+1)>$ of the logarithm of the number of solutions, shown in Figure~\ref{quenchedAverage}. Note that also this does not show any interesting behavior around $\alpha_c=4.15$.

In summary, it does not seem to be the case that the sharp-threshold phenomenon of P[UNSAT] is the simple indicator of a related, continuous phase transition of a quantitative measure.

\subsection{Are there two different phases in $k$-SAT?}
This leads us back to the question of whether we really have two phases in this system, one consisting of satisfiable instances and one consisting of unsatisfiable instances. In $k$-SAT, the main problem is that we cannot observe two different phases by eye. In this special case, the sharp threshold behavior had been observed first. and this lead to the definition of the ``phases'' instead of observing and defining the phases first before analyzing the transition between them. This happened because the sharp threshold phenomenon divided the instances into two different groups that match our intuition.  Maybe, however, an unsatisfiable instance is just an instance with $0$ solutions and not substantially different from an instance with exactly $1$ solution. The question is thus whether the two `phases' are just a differentiation that is convenient for computer scientists or whether they relate to a small structural change in some interaction on a microscopic scale that leads to a huge change in macroscopic behavior. 

\begin{figure}
\centerline{\includegraphics[width=0.9\textwidth]{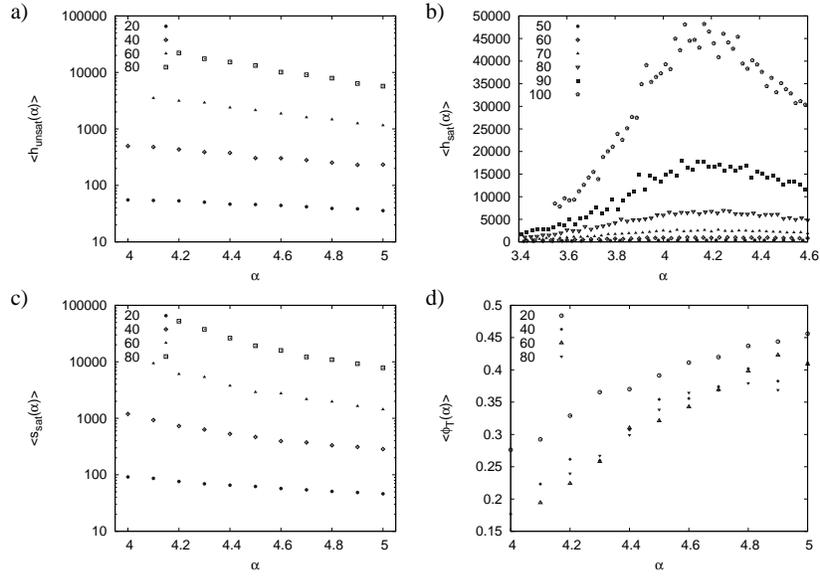}}
\caption{\label{hardness}a) Average hardness of proving that an instance is unsatisfiable, measured as the size of the decision tree in a DPL-like algorithm. b) Average hardness of finding the first solution of satisfiable instances. c) Average hardness of finding all solutions of satisfiable instances. d) Average fraction of the decision tree that is traversed until the first solution is found.}
\end{figure}

Hardness has been used to argue that there are two different phases, since it shows a diverging behavior around $\alpha_c$. Of course, hardness, measured as the number of independent decisions of a DPL-like algorithm \cite{Davis1962} or simply by the runtime, depends on the specific implementation. Nonetheless, the basic picture is always the same, namely that it peaks around $\alpha_c$\footnote{Note that the maximum itself is difficult to locate and might also shift with $n$.}. The question is whether this maximum is genuine or directly dependent on the definition of a satisfiable and an unsatisfiable instance. We will give evidence here that the occurrence of a maximal runtime around $\alpha_c$ is directly implied by the definition of a {\it decision algorithm}. The problem is that a decision algorithm does different things in the two cases: if it runs on a satisfiable instance, it stops after the first solution is encountered. Otherwise, a proof has to be given that no solution exists. For DPL-like algorithms \cite{Davis1962}, this means that in the first case only some fraction of the whole decision tree has to be searched while for unsatisfiable instances the whole tree has to be traversed. We can assume two things: 
\begin{enumerate}
\item the decision trees of typical satisfiable and typical unsatisfiable instances at a given $\alpha$ are of approximately the same size; 
\item the locations of the solutions in the leaves of the tree are uniform. 
\end{enumerate}
Thus, let the size of a typical decision tree at a given $\alpha$ be denoted by $t(\alpha)$. Even if an instance has just one solution, we will on average traverse only half of the tree to find it. For an unsatisfiable instance at the same $\alpha$, we will on average take double the time to find the solution. Since at $\alpha_c$ there are more unsatisfiable than satisfiable instances, this is already an explanation for the increasing runtime at $\alpha_c$. Of course, the behavior of the average hardness is a bit more complicated than this. The average hardness $h(\alpha)$ can be dissected into $h_{sat}(\alpha)$ and $h_{unsat}(\alpha)$, the hardness of satisfiable and unsatisfiable instances at $\alpha$. With this, 
\begin{equation}
h(\alpha) = (1-\mbox{P[UNSAT]})h_{sat}(\alpha)+ \mbox{P[UNSAT]}h_{unsat}(\alpha).
\end{equation}
Note that the hardness $h_{unsat}(\alpha)$ is simply given by the average size $t_{unsat}(\alpha)$ of the decision tree of unsatisfiable instances at $\alpha$. While $h_{unsat}(\alpha)=t_{unsat}(\alpha)$ seems to be a simple, exponentially decreasing function with $\alpha$ (s. Figure~\ref{hardness}a), $h_{sat}(\alpha)$ is at a maximum around $\alpha_c$ (s. Figure~\ref{hardness}b). $h_{sat}(\alpha)$ can be approximated as the product of $t_{sat}(\alpha)$, the size of the average decision tree of satisfiable instances at $\alpha$, and $\phi_T(\alpha)$, the average fraction of the decision tree that is traversed before a solution is found. While the first is decreasing with $\alpha$, the latter is increasing with $\alpha$. Thus, the maximum around $\alpha_c$ in $h_{sat}(\alpha)$ is introduced artificially by stopping after the first solution is encountered. 
If we instead look at the runtime of an algorithm that counts the number of {\bf all solutions} an instance has, we see no singularity of the hardness around $\alpha_c$ as Figure~\ref{hardness}. shows. We thus conclude that the hardness supports the view that there are no two phases since the size of the decision tree  decreases smoothly with growing $\alpha$, at least for the system sizes that could be computed.

Summarizing the results so far, we could not find a measure which is related to the existence question measured by P[UNSAT] and which shows a continuous phase transition. We also did not find any measure that is independent of P[UNSAT] and therefore proves that indeed an unsatisfiable instance is structurally different from an instance with $1$ solution. Instead, we will now present results from two very simple statistical systems that show a sharp threshold phenomenon. We will then use these systems to develop a simple toy model  that shows qualitatively the same behavior as $3$-SAT and shows quite clearly that no phase transition is needed to produce a $3$-SAT-like system.

\section{Sharp threshold phenomena in simple statistical systems}
\label{models}

In this section we discuss two simple stochastic processes. The first one, is a simple coin tossing example that is discussed in Sec. \ref{coin} and the second is a statistical problem, called the {\it coupon collector's problem}, discussed in Sec. \ref{couponColl}.

\subsection{Throwing a Biased Coin}
\label{coin}

\begin{figure}
\centerline{\includegraphics[width=0.9\textwidth]{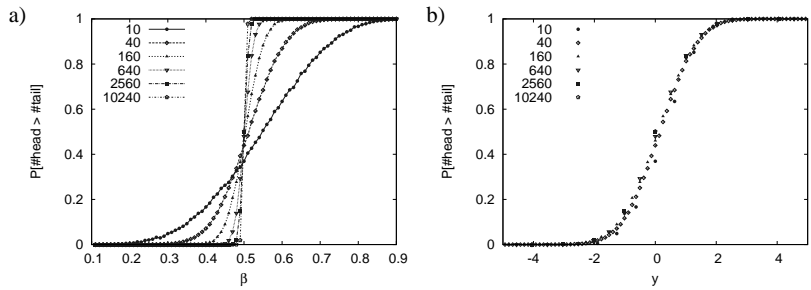}}
\caption{\label{FigCoin} a) Probability $P[\# heads > \#tails]$ that more heads than tails are tossed in $\hat{n}$ tosses with a biased coin that shows heads with probability $\beta$ and tails with probability $1-\beta$. b) $P[\# heads > \#tails]$ in dependency of the rescaled parameter $y$.}
\end{figure}

In the book {\it Computational complexity and statistical physics}, the editors briefly discuss the question of whether sharp thresholds are more than just an effect of the law of large numbers. They contrast $SAT$ with the following simple system \cite[p.8]{Percus2006}: a biased coin is tossed that shows heads with probability $\beta$ and tails with probability $1-\beta$. Let an instance consist of $\hat{n}$ tosses and let $\hat{n}$ define the system size. We expect the chance $P[\# heads > \# tails]$ to see more heads than tails in one of these instances to change from $0$ for $\beta < 0.5$ to $1$ for $\beta > 0.5$ with an ever-increasing sharpness with growing $\hat{n}$. With this example, Percus et al. indicate that sharp threshold phenomena {\it per se} are not so surprising, but they don't settle the question of whether this simple system will already show finite size scaling. 
The question is thus whether the curve $P[\#heads > \#tails]$ for low $\hat{n}$ just fluctuates stronger or is indeed less steep than that of a larger system. This question is settled by Figure~\ref{FigCoin}.

Figure~\ref{FigCoin}a shows the fraction of $10,000$ instances of $\hat{n}$ tosses each where more heads than tails were shown. The curves meet approximately at $\beta = 0.5$. Plotting them against the rescaled parameter $y = \hat{n}^{0.5}(\beta-0.5)/0.5$ shows a perfect universal scaling. This model is especially interesting since here also the sharp threshold behavior results from asking a peculiar kind of question. Instead of looking at the more natural question of $P[heads]$ which is of course identical to $\beta$, the behavior artificially becomes a sharp threshold behavior by asking when it is more likely to see more heads than tails in any given system size. Moreover, this most simple system also displays a finite size scaling effect. Naturally, the corresponding exponent $\beta = 0.5$ is the one dictated by the law of large numbers. Thus, although a finite size scaling effect can be seen, nobody would regard it as the effect of a phase transition since the exponent is a trivial one. The next example is much more interesting since it shows a non-trivial exponent.

\subsection{The Coupon Collector's Problem}
\label{couponColl}
The simple system of coin tossing cannot easily be likened to $3$-SAT. We will thus introduce a second statistical problem called the {\it coupon collector's problem}: let there be a set of $n'$ distinguishable objects called {\it coupons}, identified by a coupon ID from $1$ to $n'$. Each coupon is contained multiple times in a large multi-set and collectors can purchase coupons from this multi-set by drawing one item uniformly at random. We will assume that each coupon ID has the same probability of being drawn. The coupon collector problem asks how many draws have to be made expectedly until each coupon ID is drawn at least once, i.e., the question of when the collection is completed. In essence, once the collector has collected $k$ different IDs, the chance of picking a new ID is $\frac{n'-k}{n'}$ and thus the expected time to find a new one is $\frac{n'}{n'-k}$. Summing over these expected times gives $\frac{n'}{n'}+\frac{n'}{n'-1}+\ldots+n'=n'\left(\frac{1}{1}+\frac 1 2 + \ldots+\frac 1 {n'} \right) = n' H_{n'}$. This can be approximated to be $n' \ln {n'}+ \Upsilon n' + \frac 1 2 + O(1)$, where $\Upsilon \simeq	 0.57722$ denotes the Euler--Mascheroni constant. The variance is bound from above by $2n'^2$.

\begin{figure}
\centerline{\includegraphics[width=0.9\textwidth]{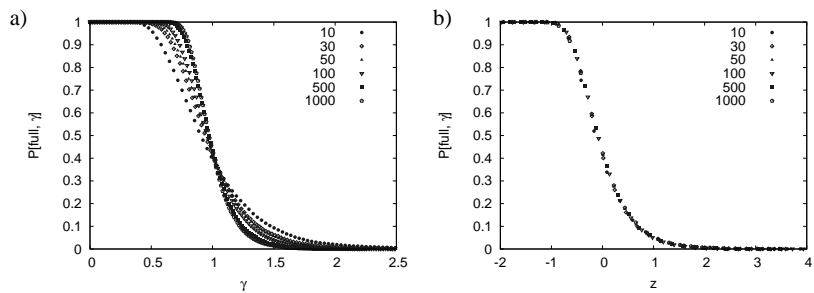}}
\caption{\label{cc} a) Percentage of full collections in dependency of $\gamma = t/(n\log n + 0.577n+0.5 )$. b) Percentage of full collections in dependency of the rescaled parameter $y$.}
\end{figure}

For a set of $x$ collections, we now define $P[full, t]$ to be the fraction of full collections after $t$ draws. Of course, the number of draws depends on the system's size. We thus define $\gamma := \frac{t}{n' \ln{n} + 0.577 n' + 0.5}$ and plot $P[full,\gamma]$ against $\gamma$.  Figure~\ref{cc}a shows the result for different system sizes from $10$ to $1000$ in dependence of $\gamma$. Interestingly, this looks like a phase transition at a critical $\gamma_c = 1$. Furthermore, we define a rescaled parameter $z = n'^{0.17}\left(\gamma-\gamma_c\right)$ against which we plot the functions, as shown in Figure~\ref{cc}b.

Note that the critical exponent is far away from the trivially expected $0.5$. We can now define two phases: full collections and incomplete collections. With this, Figure~\ref{cc} shows clearly that there exists a first-order phase transition between the two phases. Or does it?  %although this simple model cannot exhibit non-trivial collective behavior. 
But of course, a system as simple as the coupon collector's problem does not meet the intuition about a system with a phase transition and it especially cannot exhibit any non-trivial collective behavior. 
Just defining that one condition of a system, i.e., whether a collection is complete or not, represents two phases does not make them different phases. 
Also, the finite size scaling effect cannot justify the notion of a phase transition since it seems to be mainly an effect of the law of large numbers. 

In the following we will highlight the connection between the coupon collector's problem and the behavior of P[UNSAT] in $3$-SAT.

\subsubsection{Connection between random $k$-SAT and the coupon collector's problem}
When $\alpha = 0$ each random $k$-SAT instance has exactly $2^n$ solutions. Every added clause $C = \{l_1, l_2, \dots, l_k\}$ excludes all solutions in which all negated literals $l_i$ are assigned $true$ and all positive literals $l_j$ are assigned $false$. That is, each added clause extinguishes a fraction of $2^{-k}$ of all remaining solutions. Of course, some of the solutions might already have been extinguished by a clause added earlier. An instance becomes unsatisfiable when all of its possible assignments have been extinguished by some clause. Thus, the question is very similar to that of the coupon collector's problem: in each time step we draw uniformly at random $k$ literals that extinguish a $2^{-k}$th of all possible assignments and we want to know when all possible assignments are extinguished. Of course, there are two main differences: we draw more than one `coupon' at once, namely $2^{n-k}$, and moreover these are not independent of each other. The first condition alone would just reduce the expected completion time by some factor, but the effect of the second condition is harder to estimate.

Note that there is really no kind of interaction between the clauses. Given a set of solutions $S$ that are left for some instance $I$, adding a clause will lead to the following reduced set of solutions $S'$: let $s \in S$ be any solution that does not satisfy the newly added clause. This cannot be a solution of the new instance, and thus it is removed from $S$. Let now $s \in S$ be some solution that satisfies the newly added clause. Since it was contained in $S$, this means that the assignment given by $s$ satisfies at least one literal in all the clauses added so far plus at least one in the newly added clause. Thus, this solution is in $S'$. The clauses are independent of each other in the sense that the only solutions extinguished by a clause are those that don't satisfy it. There is no cumulative effect of the clauses such that after adding some of them a whole avalanche of solutions is extinguished. Note, however, that the solutions in $S$  are not independent of each other since if $s \in S$, other solutions $s'$ with a low Hamming distance to $s$ have a higher probability of being in $S$ than those with a large distance.

\subsection{A toy model for $3$-SAT}
Neither the coupon collector's problem nor coin tossing displays one of the main qualitative behaviors of $3$-SAT. The main point of interest is the gap between $\alpha = 5.19$ at which the average number of solutions meets $1$ and the point $\alpha_c$ at which most instances are already unsatisfiable. 
In the following, we introduce a toy model that shows this more involved behavior but is still quite simple and not likely to have a real phase transition. The toy model is based on the following idea: an instance is represented by a number, starting with $2^n$. This represents the number of solutions left at a given $\alpha$. Adding a clause is mimicked by multiplying this number by some reduction factor. 

Of course, simply multiplying the number by $7/8$ is already enough to produce the average number of solutions shown in Figure~\ref{avgNumberSol}, and also a sharp threshold behavior of P[UNSAT]. But, unfortunately, the latter takes place at $\alpha = 5.19$. Looking at the real reduction factor, it turns out that the distribution broadens with $\alpha$ and is shifted to the right. We used this observation for the toy model of random 3-SAT, in which we draw a multiplicative factor from a normal distribution with a standard deviation $\sigma = 0.0585*\alpha$ and an average of $\mu(\alpha)$ given by
\begin{equation}
\mu(\alpha)=\left\lbrace 
\begin{matrix}
0.875+0.009*\alpha & \alpha<3.8 \cr 
0.875+0.170*\alpha & \alpha\geq 3.8
\end{matrix}
\right. .
\end{equation}
If the drawn number is lower than 0 or higher than 1, we set it to 0 or 1, respectively. This factor is then multiplied with the current number of the toy model instance. An instance of the toy model represents an unsatisfiable instance if its number drops below 1. Thus, $P_{\mbox{toy}}[UNSAT, \alpha]$ gives the fraction of toy model instances at $\alpha$  whose number is below 1. 

\begin{figure}
\centerline{\includegraphics[width=0.9\textwidth]{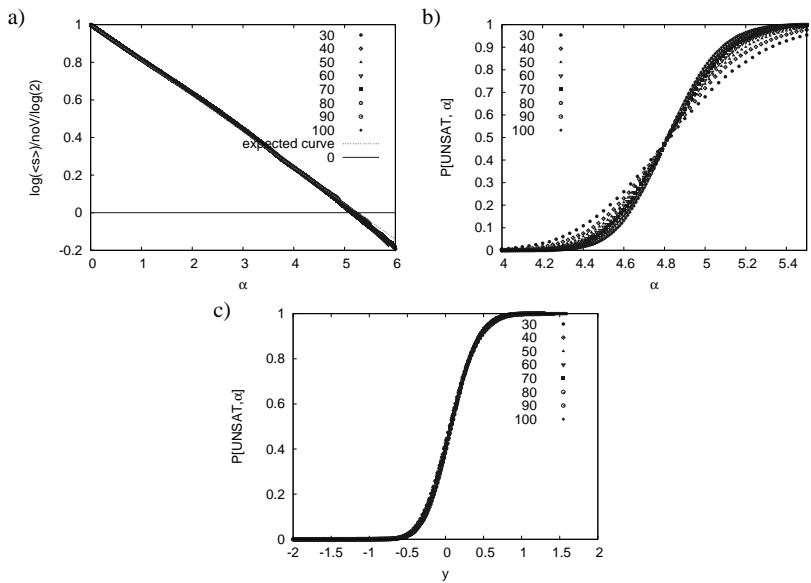}}
\caption{\label{toy} a) The average number of solutions in the toy model. As a reference, the expected curve for real $3$-SAT instances as described by equation \ref{avgNumb} is also given. b) $P_{\mbox{toy}}[UNSAT,\alpha]$ for the toy model. c) $P_{\mbox{toy}}[UNSAT,\alpha]$ against the rescaled parameter $y= N^{0.41}*(\alpha-4.76)/4.76$, i.e., the critical $\alpha_c(toy)$ is $4.76$.}
\end{figure}

\begin{figure}
\center{
\includegraphics[width = 0.6\textwidth]{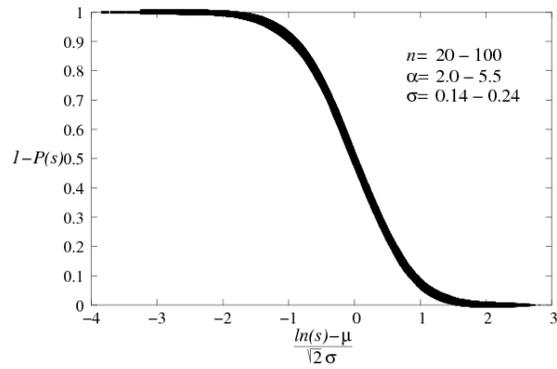}}
\caption{\label{toydist}Universal scaling of the cumulative distribution of the number of solutions in the toy model. The $\mu$ and $\sigma$ denote the two fitting parameters of the lognormal distribution used for the rescaling.}
\end{figure}

In Figures~\ref{toy} and \ref{toydist} we show our simulation results for the toy model defined above. According to Figure~\ref{toydist}a, the average number of solutions follows the same exponential behavior as expected from (\ref{avgNumb}), and $\left< s\right>$ drops below 1 at $\alpha=5.19$. Surprisingly, a sharp threshold behavior can be observed when plotting P[UNSAT] as a function of $\alpha$ as shown in Figure~\ref{toydist}b-c. Similar to $3$-SAT, the transition point of the threshold behavior at $\alpha=4.76$ is separated from the point where $\left< s\right>=1$ by a non-negligible gap. Furthermore, the distribution of the numbers $P_{\mbox{toy}}[s,\alpha ]$ is best described by a lognormal distribution and shows the same universal scaling behavior as the real $P[s,\alpha ]$ distribution, as displayed in Figure~\ref{toydist}.

In summary, this toy model shows the same qualitative properties as the real $3$-SAT system.

\section{Summary}
\label{summary}
In this article we have raised the question of whether or not the sharp threshold phenomenon displayed by P[UNSAT] around $\alpha = 4.2$ is a mere statistical event that does not relate to a phase transition in the classical sense. Our intuition is that there is no interaction of the elements of a Boolean instance, i.e., clauses, variables, or solutions, that leads to this phenomenon. We also see no principal difference between instances with at least $1$ solution and those with no solution. We thus believe that the sharp threshold behavior of P[UNSAT] can rather be likened to the sharp threshold phenomena in simpler systems, like the coupon collector's problem. Of course, it is obvious that approaches from statistical physics were successful in describing $3$-SAT and that some of these results lead to the most powerful SAT-solver based on survey propagation \cite{Monasson1999}. It is important that we not question the phase transition shown for other order parameters like  backbone size \cite{Monasson1999b}, clustering of the solution space \cite{Krakala2007}, or the order parameter associated with the messages in survey propagation \cite{Mezard2002}, but only P[UNSAT] as an order parameter of a real phase transition. 

We conclude by describing one of the possibly many examples where asking
somewhat different questions about the states of the same system may
easily lead to the conclusion that more than one observable
transition (and of different kinds)  takes place in the system,
even though it is widely accepted that there is only a single relevant
transition in it.

Consider the Ising model on a face-centered cubic lattice. As
the system cools down from high temperatures, we ask two simple
questions (without loss of generality we can assume that for low
temperatures the up spins take over):

\begin{enumerate}
\item What is the total spontaneous magnetization of the system? (ratio
of up spins minus the ratio of down spins)
\item Is there a percolating cluster of down spins present?
\end{enumerate}

The (textbook level) answers are:

\begin{enumerate}
\item  Below a critical temperature $T_c^I$,  the spontaneous magnetization
sharply increases as the number of up spins starts to grow quickly.
The associated transition is a prototype of continuous phase
transitions (involving fluctuations, etc).
\item       At a  temperature $T_p^I  <  T_c^I$,  the probability that a
percolating (connected infinite) cluster of down spins is present
suddenly drops from 1 to 0 (as if a first-order transition was taking
place).
\end{enumerate}

We suggest that the lesson from this analogy is the following: the
answer one gets depends very much on the question. Our conclusion is
that it remains to be demonstrated that asking ``What is
the probability of having a satisfiable instance in 3-SAT?'' is the right
question.  We argue that this particular question (order parameter) is not closely related
 to the variety of possible rich transitions
taking place in this paradigmatic satisfiability problem. 

Has the question of whether or not P[UNSAT] actually undergoes a phase transition, more to it than just being a simple question of how to name something? In this interdisciplinary field it is very important to be careful with terms; a phase transition is more than just a sharp threshold phenomenon and requires proof that the supposed phases behave differently in some aspect that is independent of their definition. The simple stochastical systems presented here stress the point that a sharp threshold phenomenon, even if accompanied by a non-trivial finite size scaling effect, is not enough to show a genuine phase transition - an independent proof of two different phases is needed in addition. We hope that this article will trigger a discussion about the observations to be made in categorizing a sharp threshold phenomenon as a non-trivial phase transition, and thereby support ongoing interdisciplinary research in this field.

\section*{Acknowledgement}
The authors thank A. Hartmann and G. Istrate for their numerous helpful comments on the early versions of our manuscript.
KAZ was supported by a grant by the Deutsche Akademie der Naturforscher Leopoldina (BMBF LPD 9901/8-182). This work was supported by the Hungarian National Science Fund (OTKA K68669, NK77824), the National Research and Technological
 Office (NKTH, Textrend) and the J\'anos Bolyai Research Scholarship of the Hungarian Academy of Sciences.

\newpage

\end{document}